\documentclass[prd,preprintnumbers,twocolumn]{revtex4-1}
\usepackage{mathptmx}
\usepackage[english]{babel}
\usepackage{amssymb}
\usepackage{amsfonts}
\usepackage{amsmath}
\usepackage{eucal}
\usepackage{graphicx,color}
\usepackage{epsfig}
\newcommand{\be}{\begin{equation}} \newcommand{\ee}{\end{equation}}
\newcommand{\bea}{\begin{eqnarray}} \newcommand{\eea}{\end{eqnarray}}
\newcommand{\beann}{\begin{eqnarray*}}  \newcommand{\eeann}{\end{eqnarray*}}
\newcommand{\bfig}{\begin{figure}} \newcommand{\efig}{\end{figure}}
\newcommand{\ba}{\begin{array}} \newcommand{\ea}{\end{array}}
\newcommand{\bcen}{\begin{center}} \newcommand{\ecen}{\end{center}}
\newcommand{\btab}{\begin{tabular}} \newcommand{\etab}{\end{tabular}}

\newtheorem{Proposition}{Proposition}[section]

\newtheorem{Theorem}{Theorem}[section]
\newtheorem{Lemma}{Lemma}[section]
\newtheorem{Corrolary}{Corrolary}[section]

\newcommand{\bp}{\begin{Proposition}}	\newcommand{\ep}{\end{Proposition}}
\newcommand{\bt}{\begin{Theorem}}	\newcommand{\et}{\end{Theorem}}
\newcommand{\bl}{\begin{Lemma}}		\newcommand{\el}{\end{Lemma}}
\newcommand{\bc}{\begin{Corrolary}}	\newcommand{\ec}{\end{Corrolary}}
\begin{document}

\title{Quenching the CME via the gravitational anomaly and holography}

\author{Karl Landsteiner, Esperanza Lopez, Guillermo Milans del Bosch}
\email{guillermo.milans@csic.es,\\karl.landsteiner@uam.es,\\esperanza.lopez@uam.es}
\affiliation{
Instituto de F\'{\i}sica Te\'orica UAM/CSIC, C/ Nicol\'as Cabrera 13-15, Cantoblanco, 28049 Madrid, Spain
}

\begin{abstract}
In the presence of a gravitational contribution to the chiral anomaly, the chiral magnetic effect induces an energy
current proportional to the square of the temperature in equilibrium.  In holography the thermal state corresponds to
a black hole. We numerically study holographic quenches in which a planar shell of scalar matter falls into a 
black hole and rises its temperature. During the process the momentum density (energy current) 
is conserved. The energy current has two components, a non-dissipative one induced
by the anomaly and a dissipative flow component. The dissipative component can be measured via the drag it asserts on
an additional auxiliary color charge. 
Our results indicate strong suppression very far from equilibrium.
\end{abstract}

\preprint{IFT-UAM/CSIC-088}
\maketitle
%
Anomaly induced transport phenomena have been in the focus of much research in recent years \cite{Kharzeev:2013ffa, Landsteiner:2016led}.
The prime example is the so called chiral magnetic effect (CME) \cite{Fukushima:2008xe}.
The CME has a component due to the 
gravitational contribution to the chiral anomaly.  In the presence of a magnetic field $\vec B$ and at 
temperature $T$ chiral fermions build up an energy current \cite{Vilenkin:1979ui, Landsteiner:2011cp}
\begin{equation}\label{eq:CMET}
\vec J_\epsilon = 32 \pi^2 T^2 \lambda \vec B\,.
\end{equation}
In the  background of gauge and gravitational fields chiral fermions have the anomaly
\begin{equation}\label{eq:gravanom}
\partial_\mu J^\mu =  \epsilon^{\mu\nu\rho\sigma}\left(  \alpha \, F_{\mu\nu}F_{\rho\sigma} + \lambda\, R^a_{b\mu\nu} R^b_{a\rho\sigma} \right)\,.
\end{equation}
In holography the gravitational anomaly is implemented via a mixed Chern-Simons term.
Space-time is curved in the additional holographic direction and this is what generates (\ref{eq:CMET}) 
from the mixed Chern-Simons term \cite{Landsteiner:2011iq}. Perturbative non-renormalization has been shown in
\cite{Golkar:2012kb, Hou:2012xg}. The relation of  (\ref{eq:CMET}) with the gravitational contribution to the chiral anomaly has
also been derived in a model independent manner combining hydrodynamic and geometric arguments in 
\cite{Jensen:2012kj, Jensen:2013rga}.
An additional constraint on effective actions consistent with (\ref{eq:CMET}) stems from considerations
based on global gravitational anomalies \cite{Golkar:2015oxw, Chowdhury:2016cmh}. 
Transport signatures  induced by (\ref{eq:CMET}) have recently been 
reported in the Weyl semimetal NbP \cite{Gooth:2017mbd}.

So far anomaly induced transport has mostly been studied in a near equilibrium setup in which local versions of 
temperature and chemical potentials can be defined. An application of anomaly induced transport is 
the quark gluon plasma created in heavy ion collisions \cite{Kharzeev:2015znc}.  There the magnetic field is extremely strong. It is also 
very short lived and might have already decayed before local thermal equilibrium is reached \cite{Skokov:2009qp}. 
One therefore needs a better understanding of how anomalies induce transport far out of equilibrium.
Holography is an extremely efficient tool to study both, anomalous transport phenomena and
out of equilibrium dynamics of strongly coupled quantum systems. Previous studies of anomalous transport in
holographic quenches focused on the pure $U(1)^3$ anomaly \cite{Lin:2013sga,Ammon:2016fru}. 
This motivates us to study anomalous transport induced by the gravitational anomaly in a holographic quantum quench. 

To develop intuition we first consider anomalous hydrodynamics \cite{Son:2009tf,Neiman:2010zi,Stephanov:2015roa}  in a spatially homogeneous magnetic field. 
The anomalous transport effect we want to monitor is the
generation of an energy current in the magnetic field (\ref{eq:CMET}). We start with an initial state at
temperature $T_0$ and heat the system up to a final temperature $T$.
In a relativistic theory the energy current $J^\epsilon_i = T_{0i}$ is the same as the momentum 
density $\mathcal{P}_i = T_{i0}$. Momentum is a conserved quantity and can not increase if it is
not injected into the system. On the other hand the anomaly induced energy current does increase and this increase
must be balanced by a collective flow not included in (\ref{eq:CMET}). 
The increase in the the non-dissipative anomalous energy current \ref{eq:CMET} will be exactly
counterbalanced by a dissipative contribution at any given moment.
If we introduce a uniform
but very light density of impurities carrying some additional charge drag will generate a convective current
proportional to the density of impurities \footnote{Drag in anomalous fluids has been studied in 
\cite{Rajagopal:2015roa, Stephanov:2015roa}
}. This current measures how much energy current is
generated via (\ref{eq:CMET}). We thus need to consider a system with two $U(1)$ charges. The first one has
the mixed gravitational anomaly (\ref{eq:gravanom}) and carries the magnetic flux. The second one is a anomaly free
auxiliary $U(1)$ charge that serves to monitor the build up of the current (\ref{eq:CMET}) as the temperature changes.
 
We now consider the hydrodynamics of the system. Since all spatial gradients vanish the constitutive relations are
 \begin{align}
T_{\mu\nu} &= (\epsilon + p) u_\mu u_\nu + p \eta_{\mu\nu} + \hat\sigma_B ( u_\mu B_\nu + u_\nu B_\mu) \,, \label{eq:T}\\
 J_\mu &= \rho u_\mu + \sigma_{B} B_\mu \,,\\
 J^X_\mu &= \rho_X u_\mu + \sigma_{B,X} B_\mu\,. \label{eq:BX}
\end{align} 
The specific form of the anomalous transport coefficients $\sigma_B$, $\hat\sigma_B$ and $\sigma_{B,X}$
depend on the hydrodynamic frame
choice \cite{Amado:2011zx}. In Landau frame
\begin{align}
\hat \sigma_B &=0\,\\
\sigma_{B} &= 24 \alpha \mu - \frac{\rho}{\epsilon + p} \left( 12 \alpha \mu^2 + 32 \lambda \pi^2 T^2 \right)\,\\
\sigma_{B,X} &= -\frac{\rho_X}{\epsilon + p} \left( 12 \alpha \mu^2 + 32 \lambda \pi^2 T^2 \right)\,.
\end{align}
Note that despite $J^X_\mu$ having no anomaly it
does have a non-trivial chiral magnetic transport coefficient in Landau frame. This makes the effect due to
dragging manifest.
We assume that the system is neutral with respect to the anomalous
charge, i.e. $\mu = \rho =0$. As initial condition we take $\vec{J}=\vec{J}^X = 0$ and the energy current to be given
by (\ref{eq:CMET}). Thus the energy current at any given moment is
\begin{equation}
32 \lambda \pi^2 T_0^2 \vec B = (\epsilon + p) \vec v \,,
\end{equation}
where we work in the linear response regime, i.e. we assume $\lambda B$ to be a small perturbation compared to the the 
energy density and pressure such that $u_\mu \approx (1,\vec v)$.
Solving for the velocity $\vec v$ and using the constitutive relation for the current $\vec J_X$ we find that it is given by
\begin{equation}\label{eq:curfinal}
\vec J_X = 32 \frac{\rho_X}{\epsilon + p} (T_0^2 - T^2)\pi^2  \lambda \vec B\,.
\end{equation} 
This equation determines the current build up due to drag at any given moment
if the system undergoes a slow near equilibrium time evolution such that an instantaneous temperature $T$ can be defined.
It is independent of the choice of Landau frame. For a generic conformal 
field theory $p = K T^4$ and $\epsilon = 3 p$, and we obtain
\begin{equation}\label{eq:equicurve}
 j_X = \frac{T^2/T_0^2-1}{T^4/T_0^4}\,,
\end{equation}
with $j_X\!=\! \frac{ K T_0^2 |{\vec J_X}| }{ 8 \pi^2 | \rho_X \lambda {\vec B}|}$. If a system undergoes near equilibrium evolution $j_X$ must lie for any given moment on 
that curve. Deviation from \eqref{eq:equicurve} will be a benchmark for far from equilibrium behavior.

We will now consider a holographic model that allows to implement the physics described in a non-equilibrium setting. The
action of our model is
\begin{align}\label{eq:action}
S & = \frac{1}{2\kappa^2} \int d^5x \sqrt{-g} \left( \mathcal {R} - 2\Lambda - \frac{1}{4} F^2 - \frac{1}{4 q^2} F_X^2   \right.\nonumber\\
& - \left. \frac{1}{2} ( \partial \phi)^2  + \lambda \epsilon^{MNOPQ} A_M R^A_{BNO} R^B_{APQ} \right) \,.
\end{align}
In addition to gravity and a scalar field $\phi$, it involves two gauge fields $A_M$ and $X_M$ with field strengths $F=dA$ and $F_X=dX$.
Only gauge transformations of $A$ are anomalous.
The Chern-Simons term is the holographic implementation of the mixed chiral-gravitational anomaly.
We do not include possible pure gauge Chern-Simons terms since
we want to isolate the effects of the gravitational anomaly. We set from now on $2\kappa^2=1$.

The dual field theory will be brought out of equilibrium by abruptly varying a coupling. These type of processes are known as quantum quenches \cite{Polkovnikov:2010yn}. Holographically, the leading mode of bulk fields at the AdS boundary are interpreted as a couplings for the dual field theory \cite{Ammon:2015wua,Zaanen:2015oix, Natsuume:2014sfa}. 
We will perform a quench on the coupling associated to the scalar $\phi$.
For simplicity we have chosen $\phi$ to be massless and neutral under the gauge fields $A_M$ and $X_M$. We consider the following simple boundary profile for its leading mode 
\begin{equation}\label{eq:quench}
\phi_0(t,{\vec x})= \frac{1}{2} \eta \Big(1+\tanh \frac{t}{\tau} \Big) \, ,
\end{equation}
Since \eqref{eq:action} is invariant under global shifts of $\phi$, the Hamiltonians before and after the quench will be equivalent. 
The energy density induced by the quench is a function of both its amplitude $\eta$ and time span $\tau$. 

In order to study the transport properties, it is enough to solve the equation of motion to first order both in the Chern-Simons coupling and the magnetic field. Besides, we will work in the decoupling limit of large charge $q$ for the anomaly free gauge field. Equivalently we can work at $q=1$ and treat $X^\mu$ perturbatively to first order. In that case its dynamics does not backreact onto the other sectors of the theory. With these simplifications, the equations of motion reduce to
\begin{eqnarray}
&& G_{MN}+\Lambda g_{MN} = \frac{1}{2} T_{MN}^\phi+2\lambda \epsilon_{OPQR (M}\nabla_A \big(F^{PO}R^{A\phantom{i}QR}_{\phantom{i}N)}\big) \label{eq:g} \,,\\
&& \partial_M ( \sqrt{-g} \, g^{MN} \partial_N \phi )= 0 \label{eq:scalar}\,,\\
&&\nabla_M F^{MN}=-\lambda \epsilon^{NOPQR}R^A_{\phantom{i}B OP}R^B_{\phantom{i}A QR} \label{eq:F}\,,\\
&&\nabla_M F^{MN}_X=0\,,\label{eq:FX} 
\end{eqnarray}
where $T^\phi_{MN}\!=\!\partial_M \phi \partial_N \phi -{1 \over 2} (\partial \phi)^2 g_{MN}$.

At zero order in a $\lambda$-expansion the gauge fields do not backreact on gravity and the scalar, and thus can be ignored when solving their leading dynamics.  
We parameterize the zero-order solution of the metric as 
\begin{equation}
ds^2= {1 \over z^2} \left( -f(t,z) e^{-2 \delta(t,z)} dt^2 +{dz^2 \over f(t,z)} + d{\vec x}^2 \right) \, .
\label{eq:metric}
\end{equation}
Time reparameterizations are used to fix  $\delta(t,0)\!=\!0$, such that the Minkowski metric is reproduced at the AdS boundary.

Field theories at thermal equilibrium are dual to black hole backgrounds. Our initial geometry will be a Schwarzschild black hole, for which $f\!=\!1\!-\!\pi^4 T_0^4z^4$ and $\delta\!=\!0$.
Around $t\!=\!0$ the quench takes place, creating an extra energy density on the boundary theory and equivalently, a matter shell that enters from AdS boundary into bulk.
The shell subsequently undergoes gravitational collapse and is absorbed by the original black hole. The resulting black hole represents the final equilibrium state with $T\!>\!T_0$. 
Similar holographic setups have been extensively studied in the last years \cite{Chesler:2008hg,Buchel:2012gw,Abajo-Arrastia:2014fma,Craps:2014eba}.

The gauge fields $A_m$ and $X_M$ satisfy the same equation at zero order in $\lambda$. However the different roles they should fulfill select different leading solutions.
We want $A_M$ to induce a constant background magnetic field on the boundary theory and no charge density. Choosing the magnetic field to point in the $x_3$ direction, 
this is achieved by the simple solution $F_{12}\!=\!B$ with $B$ constant throughout the bulk. Contrary we wish $X_M$ to induce a non-vanishing charge density and no boundary field strength,
which leads to 
\begin{equation}
F_{X,0z}=\rho_X z \, e^{-\delta(t,z)} \, ,
\label{eq:rox}
\end{equation}
The expectation value of its associated current is given by 
\be
J^\mu_X = \lim_{z\rightarrow 0} \sqrt{-g} \, F^{\mu z}_X \, .
\ee
The integration constant $\rho_X$ is the desired charge density.

At linear order in $\lambda$, the magnetic field induces a non-diagonal component in the metric
\begin{equation}
g_{03}={4 \lambda B \over z^2} \int_0^z z^3 e^{-\delta} \left( c +{12 \over z^2}(f-1) +T_{33}^\phi \right) \, ,
\label{eq:g03}
\end{equation}
where $c$ is an integration constant.
In the initial and final state the bulk scalar field stress tensor vanishes.
We fix the integration constant by demanding that the initial state reproduces \eqref{eq:CMET}, which implies $c\!=\!8 \pi^2 T_0^2$. In our conventions the energy current can be read off the asymptotic metric expansion as $g_{03} = {1 \over 4}T_{03} z^2 + \cdots$ \cite{deHaro:2000vlm}.

Neither the scalar field and nor anomalous gauge field receive a correction at order $\lambda$. However  the off-diagonal component of the metric backreacts on $X_M$ and generates an entry parallel to the magnetic field. It is governed by the equation
\begin{equation}
\partial_t  \Big({e^\delta \partial_t X_3 \over z f}\Big) - \partial_z \Big({e^{-\delta} f \partial_z X_3 \over z}\Big)= \rho_X \partial_z \big(z^2 g_{03}  \big)  \, .
\label{eq:X3}
\end{equation}
Although $X_3$ is sourced by the Chern-Simons term even in the initial state, 
it leads to a vanishing $J^3_X$. It is the subsequent evolution what generates a boundary current parallel to the magnetic field. We also note that due to the fact that $J^3_X$ is treated in the decoupling limit it reacts immediately to the drag.

\begin{figure}[h]
\begin{center}
\includegraphics[width=4.2cm]{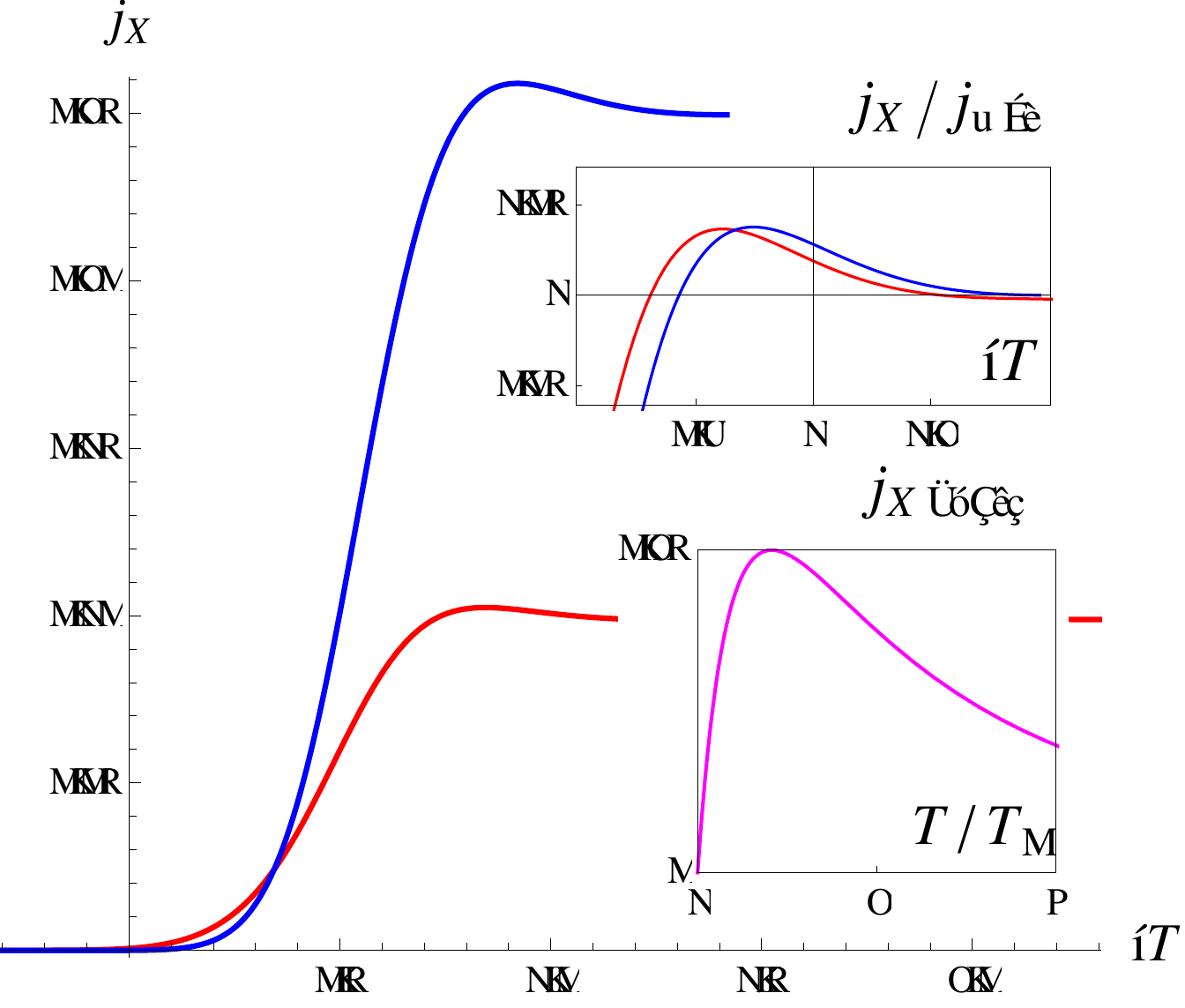}~~~
\includegraphics[width=4.2cm]{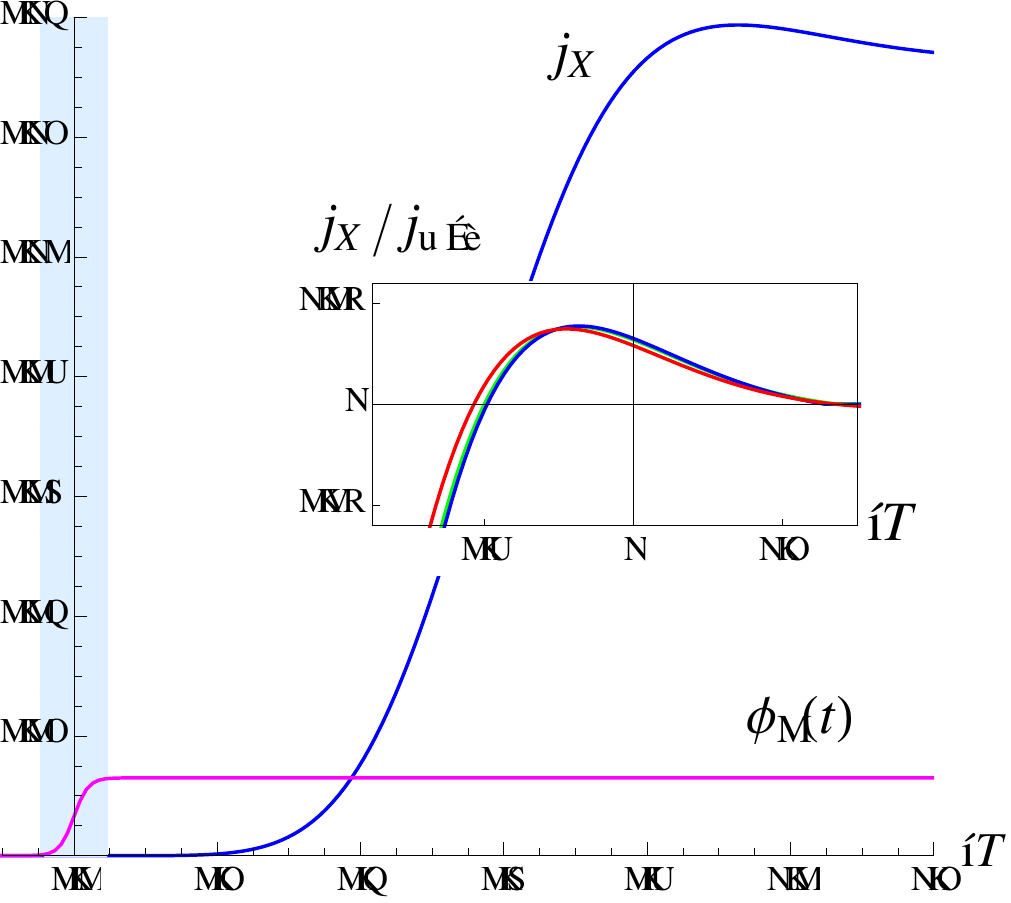}
\end{center}
\vspace{-5mm}
\caption{\label{fig:Jx} Left: Evolution of $j_X$ for quenches with $\tau T_0\!=\!0.07$ and $T/T_0\!=\!\sqrt{2}$ (blue) and $3$ (red). The final current has been normalized to one in the upper inset. The lower inset shows the equilibrium curve \eqref{eq:equicurve}. Right: Very fast quench with $\tau T_0\!=\!0.007$ and $T/T_0\!=\!2.5$. The scalar profile \eqref{eq:quench} is included, shading the time span of the quench. The inset compares this process to two more quenches with $\tau T_0\!=\!0.0175,0.035$ and same $T/T_0$.}
\end{figure}

We focus first on {\bf fast quenches}. These are processes whose time span is small in units of the inverse final temperature, $2 \tau  T\!<\!1$. 
Fig.\ref{fig:Jx}a shows the result of two such processes which the same initial temperature $T_0$ and time span $\tau$ but different final temperatures.
A comparison with the benchmark curve \eqref{eq:equicurve}, plotted in the lower inset of Fig.\ref{fig:Jx}a, rules out a possible near equilibrium description. 
In hydrodynamic process the current $j_X$ attains the maximum when the temperature reaches $T_m/T_0\!=\!\sqrt{2}$, at which it takes the value $1/4$. For higher temperatures the equilibrium current decreases, reflecting the large inertia of a hot medium. One of the processes in Fig.\ref{fig:Jx}a has been tuned to reach $T_m$, while the other generates a higher temperature $T/T_0\!=\!3$. In both cases $j_X$ exhibits a maximum before stabilizing. This contradicts \eqref{eq:equicurve}, which would predict a monotonic growth for the process whose final temperature is $T_m$, and shows that the evolution is far from equilibrium.
Moreover the current at the maximum is larger than $1/4$ in the first quench (blue) and smaller in the second (red), which again disagrees with \eqref{eq:equicurve}. 
When the currents of both processes are normalized to one at the final equilibrium state, their profiles in the rescaled time $t T$ are very similar. 
The current overshoots around $4\%$ its final value before attaining it, as can be seen in the upper inset of Fig.\ref{fig:Jx}a. 
The time span of a fast quench also has very small impact on the evolution of the current. This is illustrated in the inset of Fig.\ref{fig:Jx}b, which shows three processes with the same final temperature $T/T_0\!=\!2.5$ and different time span.

It is interesting to compare the evolution of $j_X$ with that of another important observable, the energy density. The energy density builds up during the quench and attains its final value as soon as the quench ends. In terms of the holographic model the energy pumping into the system happens while the time derivative of the scalar at the AdS boundary is non-vanishing. The profile \eqref{eq:quench} is within $3\%$ of its final value at $t/\tau\!\approx\!1.75$. We observe in Fig.\ref{fig:Jx} that instead $j_X$ equilibrates at $t T\!\approx\!1.2$. Hence the energy density and the current generated in fast quenches have independent equilibration timescales, $\tau$ and $1/T$ respectively. This provides an alternative and simple way to discard a near equilibrium evolution, since no single effective temperature can describe both observables. 
At the geometrical level the different equilibration timescales have further implications. 
While the energy density only depends on the bulk total mass, the anomaly free current must be sensitive to the interior geometry close to the emerging horizon. Consistently with this, $j_X$ turns out to mainly build up as equilibrium is approached. The body of  Fig.\ref{fig:Jx}b describes a very fast quench with $\tau T_0\!=\!0.007$. The purple curve gives the time profile of the scalar field at the asymptotic boundary \eqref{eq:quench}. 
The time interval when the quench occurs has been highlighted in blue. Notably, the current is practically zero in this example even sometime after the quench has finished.
This clearly shows that the anomaly free current does not react to the initial far from equilibrium state, but to the onset of the equilibrium. A central conclusion of our study is that anomalous transport properties related to the gravitational anomaly are very much linked to the system being at or close to thermal equilibrium.

\begin{figure}[h]
\begin{center}
\includegraphics[width=4.2cm]{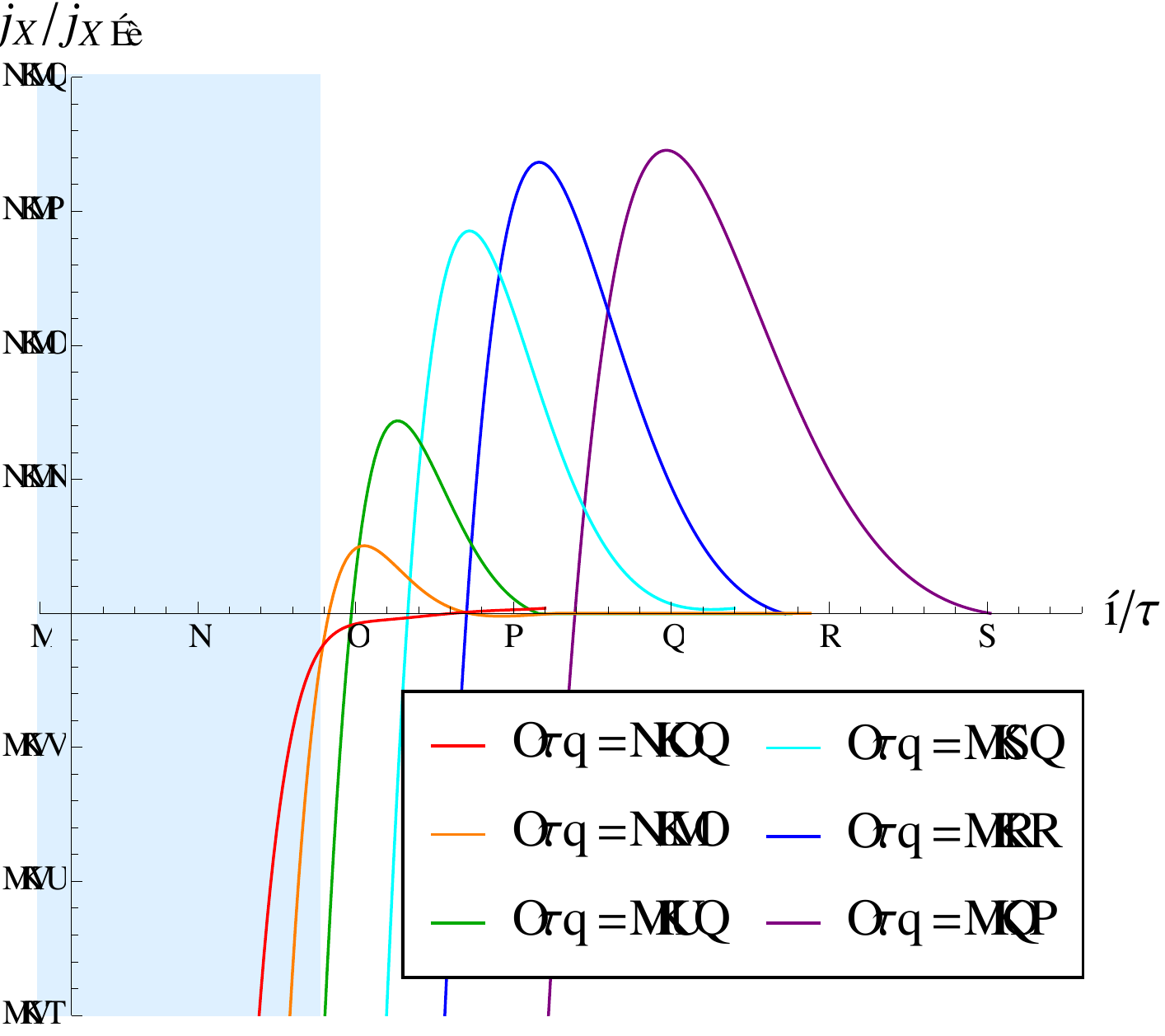}~~~
\includegraphics[width=4.2cm]{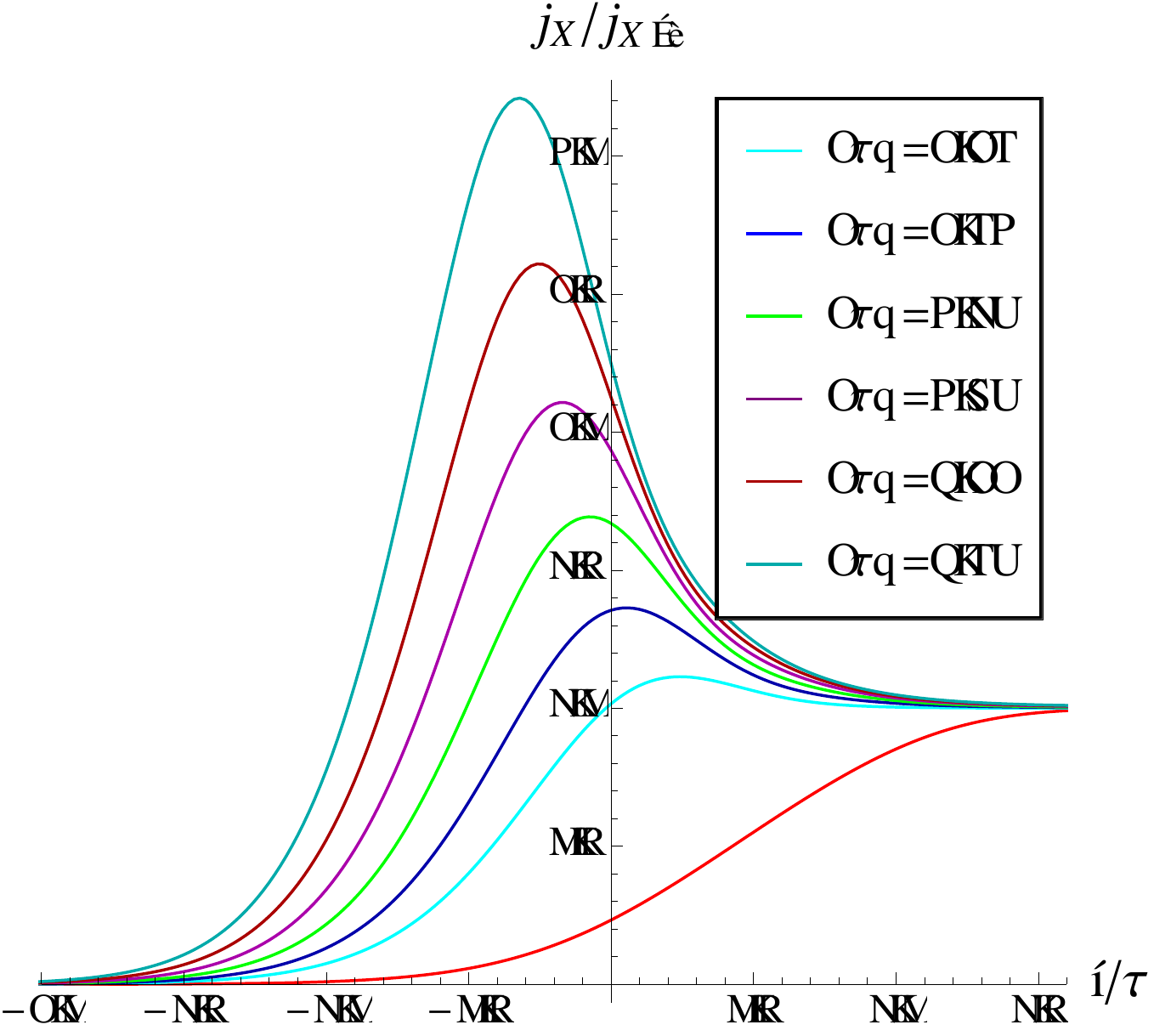}
\end{center}
\vspace{-5mm}
\caption{\label{fig:inter} Left: Final approach to equilibrium of $j_X$ for several quenches with $\tau T_0\!=\!0.175$. Right: Evolution of $j_X$ in processes with $\tau T_0\!=\!0.175$ and high final temperature. The red  curve coincides on both plots for comparison.}
\end{figure}

We wish to now study the transition from fast to slower quenches.
Fig.\ref{fig:inter}a shows the last stages in the evolution of the current for several processes with the same initial temperature and time span. 
For the sake of comparison we normalize the final current to one, and take the span of the quench instead of the final temperature as time unit.
Processes with $2 \tau T \! <\!0.5$ behave as explained above.
When $2\tau T \!\approx\! 0.65$ the maximum before equilibration starts to weaken out, and it practically disappears for $2\tau T \!\approx\! 1$. Processes with $2\tau T \!\gtrsim\! 1$ exhibit a monotonic growth of the current to its equilibrium value. We will refer to them as {\bf intermediate quenches}. We have highlighted the time span of the quench, up to the moment when $\phi_0$ is within $3\%$ of its final value. 
The transition between fast and intermediate quenches happens when time scale set by the final temperature approaches that of the quench. It is important to stress that, in general, intermediate quenches do not admit either a near equilibrium description. 
Indeed when the final temperature is higher than $T_m$, the absence of a transient maximum signals out of equilibrium dynamics. 
For the parameters associated to Fig.\ref{fig:inter}a we have $2 \tau T_m\!=\!0.5$, such that processes with final temperature below $T_m$ behave as fast quenches.

Finally we consider {\bf slow quenches}. First we consider process that are slow with respect to the final temperature and can be expected have a sizable final period of near equilibrium evolution. 
The hydrodynamic current described by the benchmark curve \eqref{eq:equicurve} decreases with rising temperature when $T\!>\!T_m$. In this case, a final period of near-equilibrium evolution implies that $j_X$ must exhibit a transient maximum. In Fig.\ref{fig:inter}b we investigate quenches with $2\tau T \!>\! 2$ for the same initial conditions as in Fig.\ref{fig:inter}a. The red curve coincides in both plots for the sake of comparison. We observe that a maximum reappears with distinctive characteristic from that exhibited by fast quenches. Contrary to them the maximum of the quotient $j_X/j_{X eq}$ increases with the temperature, and can be attained even before the quench is half way through. It is natural that the larger $\tau T$ the earlier the system enters the near-equilibrium regime, whose onset is qualitatively signaled by the maximum of the current.

\begin{figure}[h]
\begin{center}
\includegraphics[width=4.2cm]{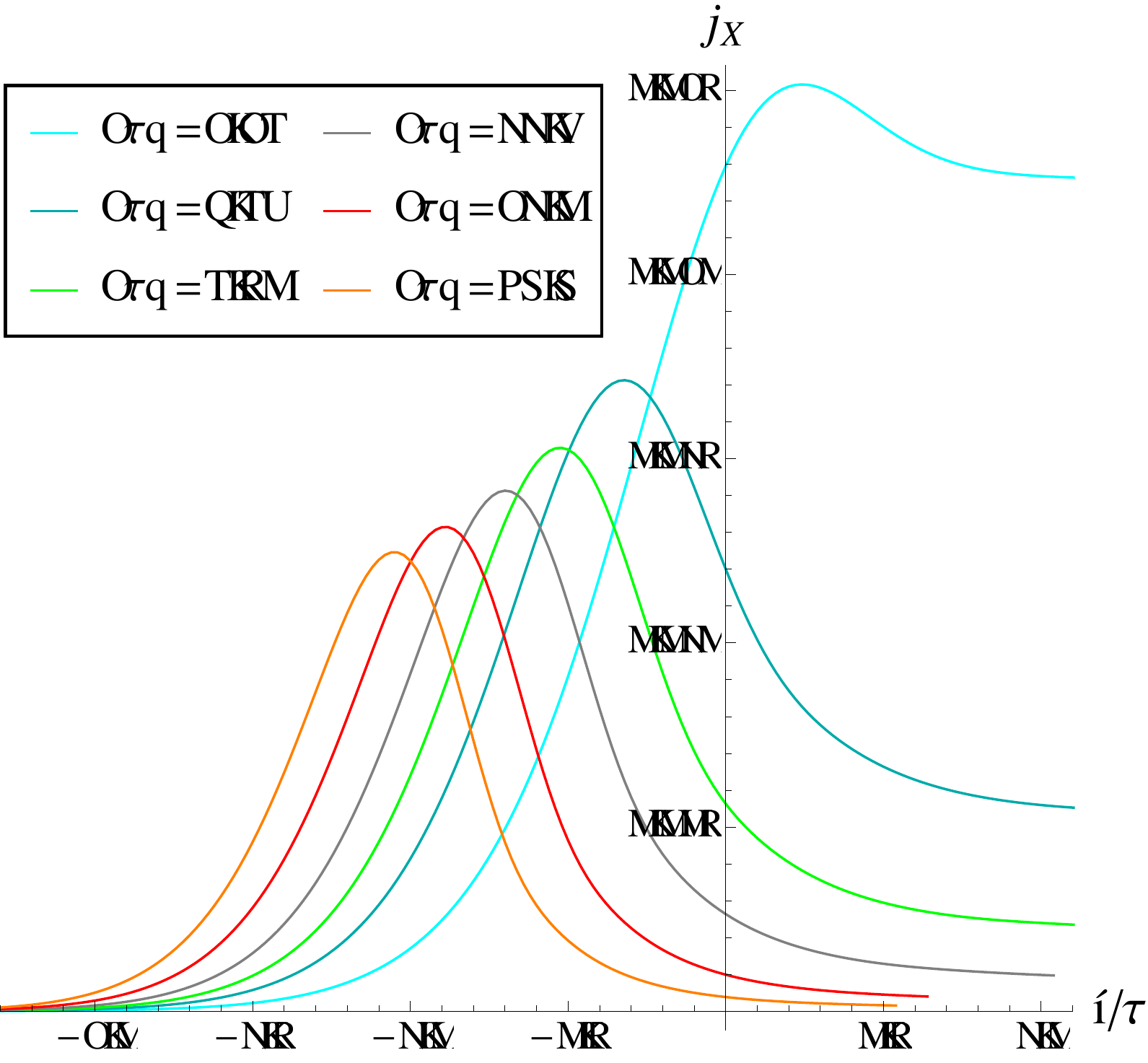}~~~
\includegraphics[width=4.2cm]{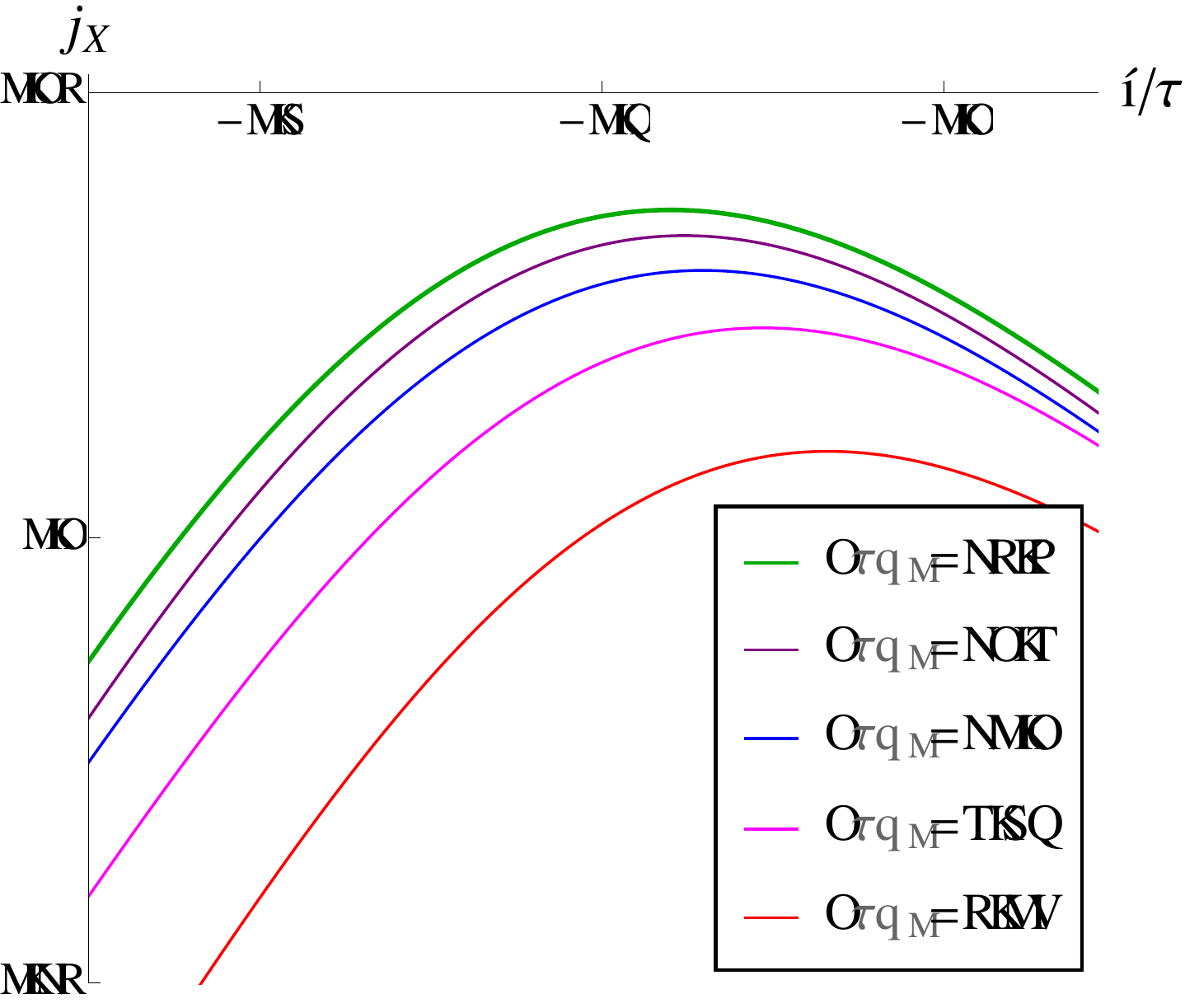}
\end{center}
\vspace{-5mm}
\caption{\label{fig:adiabatic} Left: Evolution of $j_X$ in processes with $\tau T_0\!=\!0.175$ and very high final temperature.
Right: Processes with fixed initial and final temperatures $T/T_0\!=\!2.6$ and growing $\tau$.}
\end{figure}

In the previous processes $\tau T_0\!=\!0.175$, such that they are slow with respect to the final temperature but fast with respect to the initial one. This constrains when the near equilibrium regime sets in and hence whether the current can reach $1/4$, the maximum of the equilibrium curve \eqref{eq:equicurve}. Keeping the same $\tau T_0$, Fig.\ref{fig:adiabatic}a shows evolutions with very high final temperature.
Since we are interested in the maximal value of the current, $j_X$ has not been rescaled as we did in Fig.\ref{fig:inter}.
Instead of approaching $1/4$, the maximum turns out to slowly decreases with increasing final temperature.
This shows that the near equilibrium regime in processes with $\tau T_0\!=\!0.175$ can only be associated with temperatures well above $T_m$.
In order for the complete evolution to be {\bf hydrodynamic}, the quench needs to be also slow with respect to the initial temperature. We plot in Fig.\ref{fig:adiabatic}b processes with fixed $T/T_0\!=\!2.6$ and growing time span. The maximum of the current tends now to $1/4$ as expected. We have checked that, consistently, the current builds up in a monotonic way for these
slow processes whose final temperature is equal or smaller than $T_m$.

\vspace{2mm}

We have studied holographic quantum quenches of the CME induced by the gravitational anomaly. A rich phenomenology arises depending on the time scale of the quench. 
If the quench is fast with respect to the initial and final temperatures, the evolution is far from equilibrium until the final exponential approach to stabilization. The current builds up late in the time evolution and slightly overshoots before it achieves its equilibrium value. In equilibrium the anomalous conductivity is proportional to the square of the temperature. The main motivation of this work was to analyze what activates the anomalous conductivity out of equilibrium, where there is no notion of temperature. It could have been governed by energy density, which in equilibrium is also measured by the temperature. In this case the current should have reacted as soon as energy is injected into the system. Our result on
fast quenches shows that this is not the case. Rather the system has to evolve closer to equilibrium to build up the anomalous 
current.

Intermediate quenches leave the far from equilibrium stage while they are reaching the final state, resulting in a monotonic growth of the current.
Processes which are slow with respect to the final temperature but fast with respect to the initial one, have a finite period of near-equilibrium evolution. This extends to the complete evolution for large $\tau T_0$. 

\appendix

\section{Equations of motion}

The equations of motion for leading order solution to the gravity plus scalar sector are
\begin{eqnarray}
&& \delta ' = \frac{1}{3}zT^\phi_{zz}\,, \hspace{0.72cm} f' = \frac 4z (f-1)+f\delta ' \,, \\
&& \partial_t \Big( {e^\delta \partial_t \phi \over f} \Big) = z^3 \partial_z \Big( {f e^{-\delta} \partial_z \phi \over z^3} \Big) .
\end{eqnarray}
where $f(t,z)$ and $\delta(t,z)$ are the functions appearing in the ansatz \eqref{eq:metric}. These equations are solved numerically using a fourth order Runge-Kutta algorithm.

\section{Current in initial and final state}

In equilibrium  equation \eqref{eq:g03} gives
\begin{equation}
g_{03}=\lambda B  z^2 (c - 8 \pi^4T^4 z^2) \,. \label{g03}
\end{equation}
The energy current is read off the expansion $g_{03} =\frac 1 4 T_{03} z^2 + \cdots$
\begin{equation}
J_\epsilon=T_{03}=4\lambda B c.
\end{equation}
We set $c=8\pi^2T_0^2$  such that our initial current is given by (\ref{eq:CMET}). Furthermore we note that for this value of $c$, $g_{03}$ vanishes at the horizon at the initial temperature $T_0$. 

Equation \eqref{eq:X3} for the anomaly free gauge field reduces to
\begin{equation}
\partial_z X_3=- \rho_X z f^{-1} (z^2 g_{03} +d) \, ,
\end{equation}
with $d$ another integration constant. 
Requiring $X_3$ to be regular at the horizon imposes $d\!=\!0$ in the initial state,
which turns out to yield a vanishing $J_X^3$. 
Regularity of $X_3$ is preserved by the time evolution.
Therefore in the final state with $T>T_0$ we find $d=8\lambda B\pi^{-2}(T_0^2-T^2)/T^4$. This leads
to the current
\begin{equation}
J_X =\rho_X \frac{8\lambda B}{\pi^2T^4}(T_0^2-T^2) \, .
\end{equation}
To recover equation (\ref{eq:curfinal}), note that the energy density can be read off the expansion of the metric when put in Fefferman-Graham coordinates as  $\epsilon=3\pi^4T^4$ and that $\varepsilon=3p$.

\acknowledgments{
We want to thank F. Pena-Benitez for discussions. The work has been supported by Severo Ochoa Programme grant SEV-2016-0597 and by FPA2015-65480-P (MINECO/FEDER).
}

\end{document}